\documentstyle[psfig]{mn}

\newif\ifAMStwofonts
\ifoldfss
  \ifCUPmtlplainloaded \else
    \NewTextAlphabet{textbfit} {cmbxti10} {}
    \NewTextAlphabet{textbfss} {cmssbx10} {}
    \NewMathAlphabet{mathbfit} {cmbxti10} {} % for math mode
    \NewMathAlphabet{mathbfss} {cmssbx10} {} %  "   "    "
  \fi
  \ifAMStwofonts
    \ifCUPmtlplainloaded \else
      \NewSymbolFont{upmath} {eurm10}
      \NewSymbolFont{AMSa} {msam10}
      \NewMathSymbol{\upi}     {0}{upmath}{19}
      \NewMathSymbol{\umu}     {0}{upmath}{16}
      \NewMathSymbol{\upartial}{0}{upmath}{40}
      \NewMathSymbol{\leqslant}{3}{AMSa}{36}
      \NewMathSymbol{\geqslant}{3}{AMSa}{3E}

    \fi
  \fi
\fi % End of OFSS

\ifnfssone
  \newmathalphabet{\mathit}
  \addtoversion{normal}{\mathit}{cmr}{m}{it}
  \addtoversion{bold}{\mathit}{cmr}{bx}{it}
  \newmathalphabet{\mathbfit} % math mode version of \textbfit{..}
  \addtoversion{normal}{\mathbfit}{cmr}{bx}{it}
  \addtoversion{bold}{\mathbfit}{cmr}{bx}{it}
  \newmathalphabet{\mathbfss} % math mode version of \textbfss{..}
  \addtoversion{normal}{\mathbfss}{cmss}{bx}{n}
  \addtoversion{bold}{\mathbfss}{cmss}{bx}{n}
  \ifAMStwofonts
    \ifCUPmtlplainloaded \else
      \UseAMStwoboldmath
      \makeatletter
      \new@mathgroup\upmath@group
      \define@mathgroup\mv@normal\upmath@group{eur}{m}{n}
      \define@mathgroup\mv@bold\upmath@group{eur}{b}{n}
      \edef\UPM{\hexnumber\upmath@group}
      \new@mathgroup\amsa@group
      \define@mathgroup\mv@normal\amsa@group{msa}{m}{n}
      \define@mathgroup\mv@bold\amsa@group{msa}{m}{n}
      \edef\AMSa{\hexnumber\amsa@group}
      \makeatother
      \mathchardef\upi="0\UPM19
      \mathchardef\umu="0\UPM16
      \mathchardef\upartial="0\UPM40
      \mathchardef\leqslant="3\AMSa36
      \mathchardef\geqslant="3\AMSa3E
    \fi
  \fi
\fi % End of NFSS release 1

\ifnfsstwo
  \DeclareMathAlphabet{\mathbfit}{OT1}{cmr}{bx}{it}
  \SetMathAlphabet\mathbfit{bold}{OT1}{cmr}{bx}{it}
  \DeclareMathAlphabet{\mathbfss}{OT1}{cmss}{bx}{n}
  \SetMathAlphabet\mathbfss{bold}{OT1}{cmss}{bx}{n}
  \ifAMStwofonts
    \ifCUPmtlplainloaded \else
      \DeclareSymbolFont{UPM}{U}{eur}{m}{n}
      \SetSymbolFont{UPM}{bold}{U}{eur}{b}{n}
      \DeclareSymbolFont{AMSa}{U}{msa}{m}{n}
      \DeclareMathSymbol{\upi}{0}{UPM}{"19}
      \DeclareMathSymbol{\umu}{0}{UPM}{"16}
      \DeclareMathSymbol{\upartial}{0}{UPM}{"40}
      \DeclareMathSymbol{\leqslant}{3}{AMSa}{"36}
      \DeclareMathSymbol{\geqslant}{3}{AMSa}{"3E}
    \fi
  \fi
\fi % End of NFSS release 2

\ifCUPmtlplainloaded \else
  \ifAMStwofonts \else % If no AMS fonts
    \def\upi{\pi}
    \def\umu{\mu}
    \def\upartial{\partial}
  \fi
\fi

\title{A photometric investigation of the young open 
cluster Trumpler~15\thanks{Based on observations carried out at ESO La Silla; Data
are available at the WEBDA site: http://obswww.unige.ch/webda/}}
\author[Giovanni Carraro]
       {Giovanni Carraro\thanks{e-mail: giovanni.carraro@unipd.it}\\ 
        Dipartimento di Astronomia, Universit\`a di Padova,
	Vicolo Osservatorio 2, I-35122, Padova, Italy
}
\date{Accepted.
      Received;
      in original form}

\pagerange{\pageref{firstpage}--\pageref{lastpage}}
\pubyear{2001}

\begin{document}

\maketitle

\label{firstpage}

\begin{abstract}
In this paper we  present and analyze new CCD $UBVRI$ photometry 
down to $V~\approx$~21 in the region of the 
young open cluster Trumpler~15, located in the Carina spiral feature.
The cluster is rather compact and has a core radius of about 2$^{\prime}$, which translates
in about 1 pc at the distance of the cluster.
We provide the first CCD investigation and update its fundamental parameters.
We identify 90 candidate photometric members on the base of the position in
the color-color and  color-magnitude diagrams.
This sample allows us to obtain
a distance of 2.4$\pm$0.3 kpc from the Sun and a reddening 
E$(B-V)$~=0.52$\pm0.07$.
We confirm that the cluster is young, and fix
a upper limit of 6 million yrs to its age .\\ 
In addition, we draw the attention on the lower part of the Main
Sequence (MS) suggesting that some stars can be in contracting phase and
on a gap in the MS, that we show to be a real feature,
the $B1-B5$ gap found in other young open clusters.\\
We finally study in details the extinction toward Trumpler~15 concluding
that it is normal and suggesting a value of 2.89$\pm$0.19 for the ratio
of total to selective absorption $R_V$.\\
\end{abstract}

\begin{keywords}
Open clusters and association--Individual: Trumpler~15
\end{keywords}

\section{Introduction}
In this paper we study the stellar content of the young 
compact open
cluster Trumpler~15 by means of deep multicolor CCD
photometry.\\
This  open cluster  ($\alpha$~=~10:44:33.0, $\delta$~=~ -59:24:24.0,
$l$~=~287.41, $b$~=~ -0.41; J2000.0) is located 
near the northern edge of the Great Carina Nebula (NGC~3372),
about $20^{\prime}$ above $\eta$~Carin\ae~. It
is also named VdB-Hagen~104, Lund~558 and OCL~825.\\
Like other young clusters in this region (e.g. Trumpler~14 and 16),
is rather compact and rich.
There are several intriguing questions related to this cluster,
which was discussed in the past often leading to contradictory results.
Is the interstellar
extinction toward Trumpler~15 normal?
Is this cluster connected with the other ones located much
closer to $\eta$~Carin\ae~, like Trumpler~14, 16 and Collinder~232?
In other words, does it share the same properties
of these clusters, like age and distance,
suggesting that it probably formed together with them
 in the same Star Formation event?\\
\noindent
Aiming at clarifying these issues and deriving updated estimates for its 
fundamental parameters, like distance and  age,
in this paper we 
present and discuss the first $UBVRI$ CCD photometric study
 of Trumpler~15.\\

The layout of the paper is as follows: Section~2 presents briefly
the data acquisition and reduction. In Section~3 we discuss
previous investigations on this cluster; in Section~4 we compare
our photometry with previous ones and present our data.
Section 5 illustrates the technique to derive
reddening and membership
of stars in  Trumpler~15. Section~6 is dedicated
to the study of the interstellar extinction toward the cluster,
while in Section~7 we derive estimates for Trumpler~15 age and 
distance. Finally, Section~8 discusses the geometrical structure
of the cluster and Section~9 summarizes
our findings.

\begin{figure*}
\centerline{\psfig{file=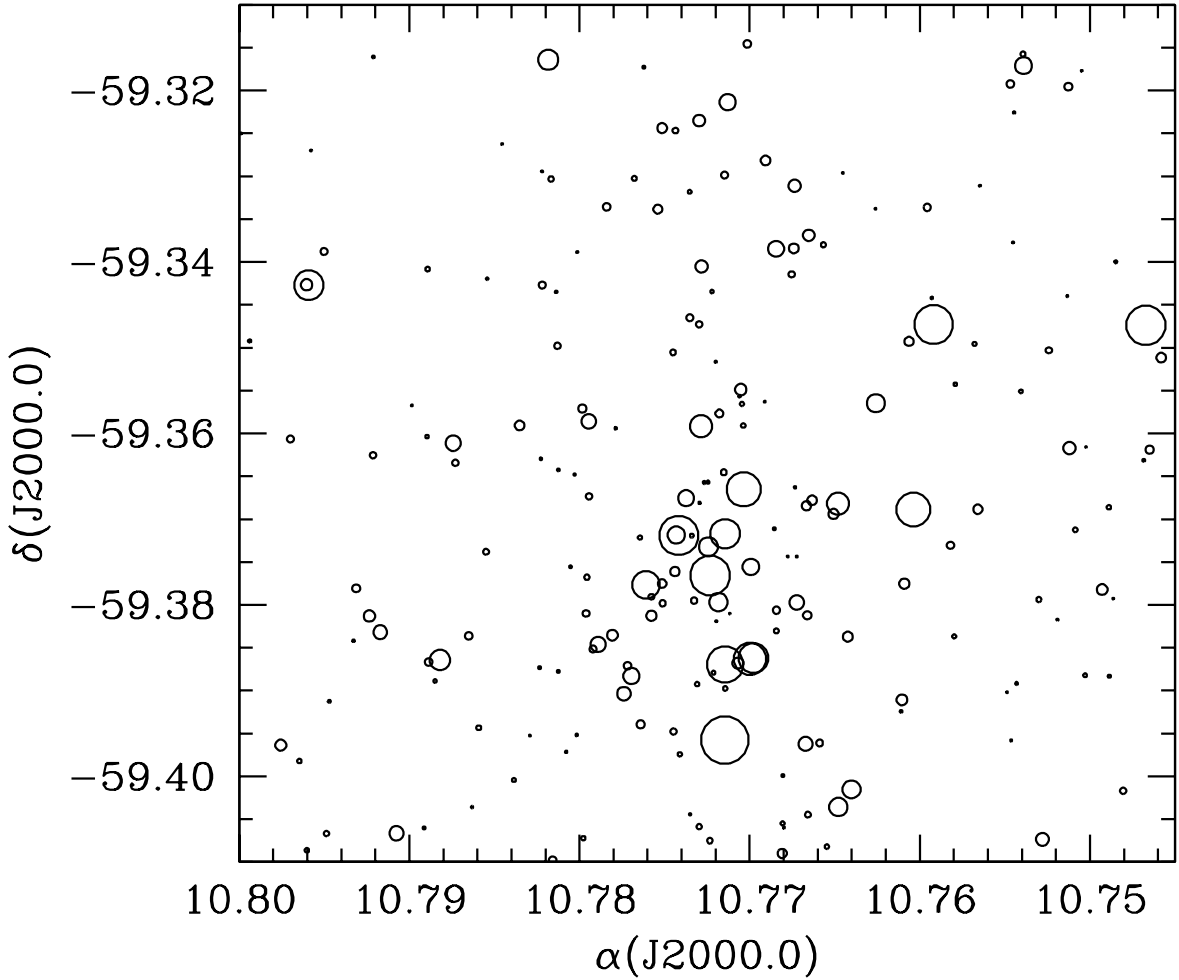,width=17cm,height=17cm}}
\caption{A map of a observed region around Trumpler~15.
The size of each star is proportional to its magnitude. North
is up, East on the left. The field is about $6^{\prime} \times 
6^{\prime}$.}
\end{figure*}

\section{Observations and Data Reduction}
Observations of four $3^{\prime}.2 \times 3^{\prime}.2$ 
overlapping fields in the region of Trumpler~15
were conducted at La Silla on April 16, 1996,
with the 0.92m ESO--Dutch telescope. 
The observations strategy, the data
reduction and  the error analysis 
have been presented in Patat \& Carraro (2001), which the reader is
referred to for any detail. 
Briefly,
the night was photometric with an average seeing of 1.0 arcsec.
In the same night we observed NGC~3324, a young open cluster
in the same region, about which we reported
elsewhere (Carraro et al 2001).
The log-book of the observations reporting the filters used,
the integration times and the seeing of each frame is given in Table~1,
whereas a map of the covered region is shown in Fig.~1.
We performed Point Spread Function (PSF) photometry and
corrected stellar magnitudes for aperture corrections, since
standard stars were measured by performing aperture photometry
(Patat \& Carraro 2001).
Our photometry is rather accurate: typical global RMS errors are 
0.03, 0.04 and 0.07 at $V$~= 12, 20 and 21, 
respectively (see Patat \& Carraro 2001).\\
The complete photometry of Trumpler~15 will be  available at the WEBDA site\footnote{
http://obsww.unige.ch/webda/navigation.html}.

\begin{table}
\tabcolsep 0.30truecm
\caption{Journal of observations of Trumpler~15 (April 16 , 1996)}
\begin{tabular}{cccc} 
\hline
\multicolumn{1}{c}{Field}    &     
\multicolumn{1}{c}{Filter}    &     
\multicolumn{1}{c}{Time integration}&         
\multicolumn{1}{c}{Seeing}         \\
      &        & (sec)     & ($\prime\prime$)\\

\hline  
 $\#1$   &   &      &      \\
	 & U &   30 &  1.2 \\
	 & B &   10 &  1.2 \\
	 & B &  300 &  1.0 \\
	 & V &    3 &  1.1 \\
	 & V &  120 &  1.1 \\
	 & R &    1 &  1.0 \\
	 & R &   60 &  1.0 \\
	 & I &    3 &  1.1 \\
	 & I &  120 &  1.0 \\
 $\#2$   &   &      &      \\
	 & U &   30 &  1.0 \\
	 & B &   10 &  0.9 \\
	 & B &  300 &  0.9 \\
	 & V &    3 &  0.9 \\
	 & V &  120 &  0.8 \\
	 & R &    1 &  0.8 \\
	 & R &   60 &  0.9 \\
	 & I &    3 &  1.0 \\
	 & I &  120 &  1.1 \\ 
 $\#3$   &   &      &      \\
	 & U &   20 &  1.0 \\
	 & B &    3 &  1.1 \\
	 & B &  300 &  1.1 \\
	 & V &    3 &  1.0 \\
	 & V &  120 &  1.0 \\
	 & R &    1 &  1.0 \\
	 & R &   60 &  1.0 \\
	 & I &    3 &  1.0 \\
	 & I &  120 &  1.1 \\
 $\#4$   &   &      &      \\
	 & U &   20 &  1.1 \\
	 & B &    3 &  1.0 \\
	 & B &  300 &  1.0 \\
	 & V &    3 &  1.0 \\
	 & V &  120 &  1.0 \\
	 & R &    1 &  1.0 \\
	 & R &   63 &  1.1 \\
	 & I &    5 &  1.1 \\
	 & I &  120 &  1.1 \\ 
\hline
\end{tabular}
\end{table}

\section{Previous studies}

\subsection{Photometric investigations}
Trumpler~15 was firstly studied by Grubissich (1968) and Th\'e \& Vleeming (1971).
These latter provided $UBV$ photographic photometry for 25 stars down to 
$V$~=~13,
confirming previous suggestions by  Grubissich (1968) that the cluster
is very young. Moreover they find that the cluster is closer to us than
Trumpler~14 and 16, at about 1.6 kpc.\\
A more detailed study has been carried out by Feinstein et al (1980),
who measure 48 stars in $UBVRI$
using an RCA 1P21 photo-multiplier down to $V$~=~14.
They identify 24 candidate members, and suggest that probably other 12 stars
can be members of the cluster.\\
While they basically confirm the cluster age, they significantly revise
the distance, putting Trumpler~15 at 2.6$\pm$0.2 kpc, roughly at the same
distance of Trumpler~14 and 16. From the analysis of their photometry
they conclude that the reddening law in the direction of the cluster is normal
and that the  mean color excess is E$(B-V)~=~0.48\pm0.07$.\\
Trumpler~15 was also observed in the Walraven photometric system by
Th\'e et al (1980) who found a distance of 2.5 kpc and a very low reddening 
E$(B-V)$~=~0.189.\\
Finally, 
Tapia et al (1988) present $JHKL$ near infra-red photometry 
for 35 stars and, at odds
with  Feinstein et al (1980), claim that the interstellar extinction
toward Trumpler~15  is not normal and report  a ratio of total
to selective absorption $R_V ~=~4.0$, significantly higher than the widely
accepted  value of 3.1, which instead seems to 
hold for all the other clusters
in the vicinity of $\eta$~Carin\ae~ (Feinstein et al 1973).\\

In conclusion there seems to exist a general disagreement in all the major properties
of Trumpler~15, but for the age.

\subsection{Spectroscopic investigations}
Photographic spectroscopy of 21 stars in the field of Trumpler~15 has been
carried out by Morrell et al (1988). This investigation led to the spectral
classification of the studied stars, all with magnitude brighter than $B$~=13.
There stars were found to range from $O$ to $B9$ spectral type. For some of these
stars the classification were already been obtained by Feinstein et al(1980),
and the two sources compare very nicely.

\begin{figure}
\centerline{\psfig{file=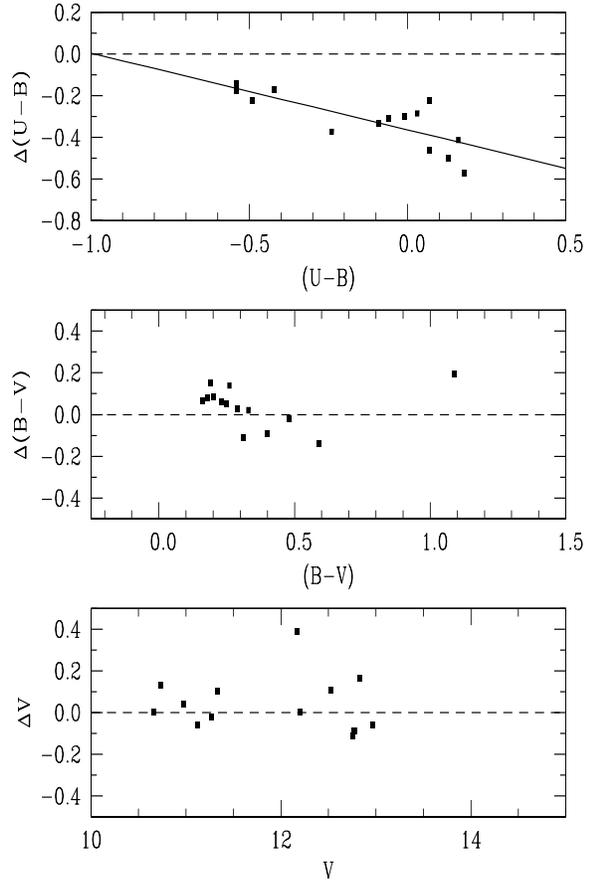,width=9cm,height=13cm}}
\caption{A comparison of our photometry with Th\'e \& Vleeming
(1971) study. The comparison is in the sense (this study - Th\'e \& Vleeming).}
\end{figure}

\begin{figure}
\centerline{\psfig{file=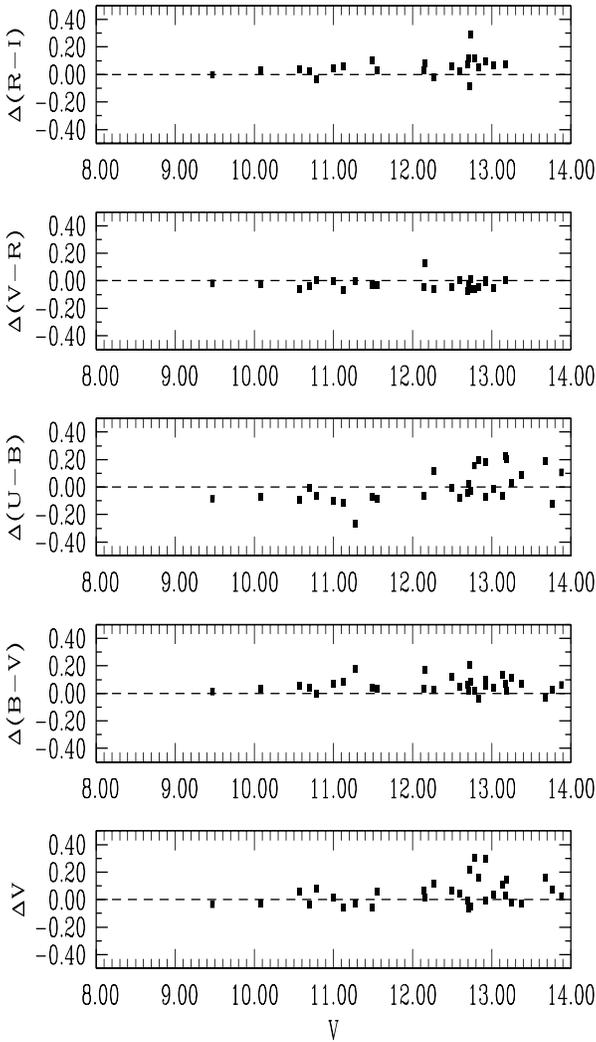,width=9cm,height=15cm}}
\caption{A comparison of our photometry with Feinstein et al
(1980) study. The comparison is in the sense (this study - Feinstein et al).}
\end{figure}

\section{The present study}
We provide $UBVRI$ photometry for 1404 stars in a $6^{\prime}.2 \times
6^{\prime}.2$ region centered in Trumpler~15, up  to about $V~=~21$.
The region we sampled is shown in Fig.~1, where a
$V$ map is presented. In this map North is on the top, East on
the left. The stars are plotted according to their magnitude.
Open symbols are used to illustrate how blending situations
are actually present.\\
Fig.~2 and 3 present a comparison between our photometry and 
 Th\'e \& Vleeming (1971) and Feinstein et al (1980), respectively.\\
A comparison with Th\'e \& Vleeming (1971) photometry is shown in Fig.~2
for 16 stars in common. From this sample we find

\[
V_{C} - V_{TV} = 0.08\pm0.18
\]
 
\[
(B-V)_{C} - (B-V)_{TV} = -0.01\pm0.09
\]

\[
(U-B)_{C} - (U-B)_{TV} = -0.28\pm0.13   ,
\]

\noindent
where the suffix $C$ stands for the present work, while $TV$ stands for
Th\'e \& Vleeming (1971). The difference is large only
for $(U-B)$, but  the scatter is large in all the relations, 
as can be noticed also by inspecting Fig.~2.
It is important to stress that the field is rather crowded and
unevenly obscured, and usually
photographic photometry performs badly at the faint magnitude end and
in cases of blended stars. Th\'e \& Vleeming (1971) point out  that
they were not able to measure the three brightest stars in the cluster core
(see Fig.~1) due to blending.\\
The larger difference is in the $(U-B)$ color. Such discrepancy  is
reported also by Feinstein et al (1980), 
who in  fact report differences with 
Th\'e \& Vleeming (1971) of 0.01$\pm$0.12, 0.07$\pm$0.15, and 0.42$\pm$0.16
for $V$, $(B-V)$  and $(U-B)$, respectively. Although for  $V$ and
$(B-V)$ there is a good agreement, nonetheless the scatter is quite large.
Apparently, a  calibration problem with $(U-B)$ exists
in Th\'e \& Vleeming (1971), and most probably the
reason of the scatter is an unaccounted color term in the photographic
measurements, as the solid line in
Fig.~2 (upper panel) would seem to indicate.\\

We have compared our photometry with Feinstein et al (1980) for
33 common stars. The results are shown in Fig.~3. We obtain

\[
V_{C} - V_{FFM} = 0.03\pm0.08
\]
 
\[
(B-V)_{C} - (B-V)_{FFM} = 0.05\pm0.09
\]

\[
(U-B)_{C} - (U-B)_{FFM} = 0.06\pm0.15
\]

\[
(V-R)_{C} - (V-R)_{FFM} = -0.03\pm0.05
\]

\[
(R-I)_{C} - (R-I)_{FFM} = +0.02\pm0.06
\]

\noindent
where the suffix $FFM$ refers to Feinstein et al (1980).
It appears that the two photometries are basically
consistent within the errors.
However, the scatter is significant.
We already reported in other studies (Carraro et al 2001, Patat \& Carraro
2001) significant scatter when comparing 
photoelectric and CCD photometry,
especially for the cases of young compact clusters
like NGC~3324 and Bochum~11.
The reasons for such scatter can be blending effects, 
stars variability, uneven obscuration and so forth. We believe
that the main source of error is the difficulty of photoelectric
aperture photometry in dealing with crowded fields. For instance in the case
of Trumpler~15, we found that the stars \#14 (Grubissich numbering) is
actually a blend of two stars that thanks to the good seeing conditions
we were able to separate. Another case for which the difference is huge
is stars \#11, which has a close poorly resolved companion.\\

Finally, we have compared our photometry with the Walraven photometry from
Th\`e et al (1980). To perform the comparison, we translated the 
Walraven $(V-B)$ color
into the Johnson $(B-V)$ color by using the relation

\begin{equation}
(B-V)_J=2.571 \cdot (V-B)-1.020 \cdot (V-B)^2 +
0.500 \cdot (V-B)^3 -0.01
\end{equation}

\noindent
which is valid for $(V-B)$ smaller than 0.45.
By considering all the stars in common (22) we obtain:

\[
(B-V)_{TBA} - (B-V)_C = -0.048 \pm 0.075
\]

\noindent
However, four stars - Grubissich (1968) numbers 2, 6, 8 and 22 deviate 
significantly (see also Feinsten et al 1980).
By excluding these stars, we obtain

\[
(B-V)_{TBA} - (B-V)_C = -0.037 \pm 0.038
\]

\noindent
which indicates that the two photometries are consistent within the errors.
From the comparison we have excluded stars $\#14$, that we have already
shown to be a blend of two stars.\\

\noindent
The CMDs for all the measured stars is plotted in Fig.~4 in the planes
$V-(B-V)$, $V-(V-I)$ and $V-(V-R)$.
The dashed line indicates  the magnitude limit of Feinstein et al (1980)
photoelectric photometry, the deepest one before the present study.\\
Our photometry reaches $V\approx 21$, and
the Main Sequence (MS) extends for about 10 mag.
It gets wider at increasing magnitude, due to :
\begin{itemize}
\item some -but not severe- field star contamination;
\item photometric errors, which typically amount at $\Delta (B-V)$~=~
0.04, 0.10 and 0.15 at $V$~=~12.0, 16.0 and 20, respectively.
\item the presence of unresolved binary systems, whose 
percentage in these clusters is around 30$\%$ (Levato et al 1990).
\end{itemize}

\noindent
Two interesting features are worth to be noticed:
\begin{description}
\item 1) the presence of a prominent gap at $V \approx 12$, 
$(B-V) \approx 0.2$;
\item 2) a change of slope in the MS at $V \approx 14.5$, which is a
hint for a possible {\it turn on}.
\end{description}

\begin{figure}
\centerline{\psfig{file=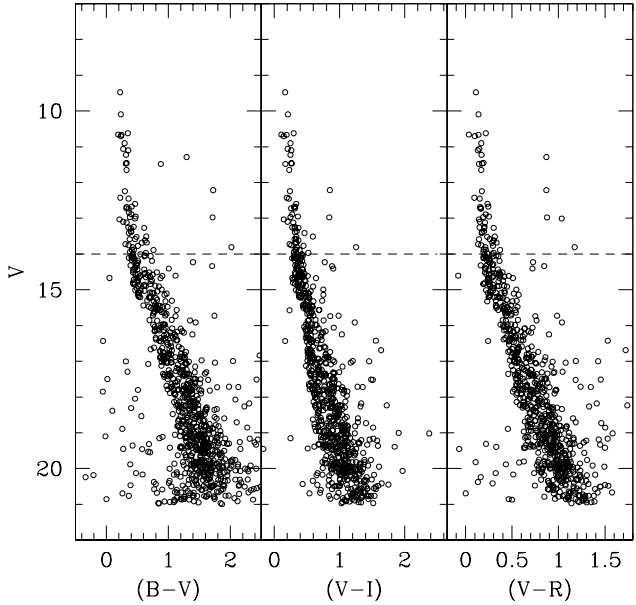,width=9cm,height=9cm}}
\caption{The CMDs of Trumpler~15 including all the detected stars.}
\end{figure}
 
\section{Individual reddenings, membership and differential 
reddening}
We use $UBV$ photometry to derive stars' individual
reddenings and membership, and the analysis strictly follows
Carraro \& Patat (2001) and Patat \& Carraro (2001).
Basically we determine individual reddenings by means
of the reddening free parameter $Q$ and the distribution
of the stars in the $(U-B)-(B-V)$ diagram.\\

\subsection{Individual reddening and membership}
The color-color diagram for all the stars having $UBV$ photometry
(142 stars in total) is shown in Fig.~5.
In this plot the solid line is an empirical ZAMS taken
from Schmidt-Kaler (1982). \\

\begin{figure}
\centerline{\psfig{file=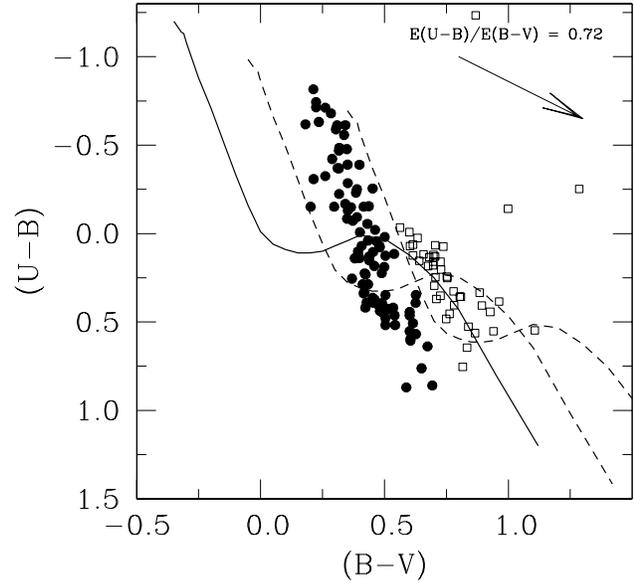,width=9cm,height=9cm}}
\caption{Two color diagram for stars in the field of Trumpler~15
having $UBV$ photometry. Filled circles indicate stars having mean
reddening $E(B-V)~=~0.52\pm0.07$ (r.m.s.), open squares stars with larger reddening
The solid line is the empirical ZAMS
from Schmidt-Kaler (1982), whereas the dashed lines
are the same ZAMS, but shifted by E$(B-V)$~=~0.30 and by E$(B-V)$~=~0.70, 
respectively.}
\end{figure}

\begin{figure}
\centerline{\psfig{file=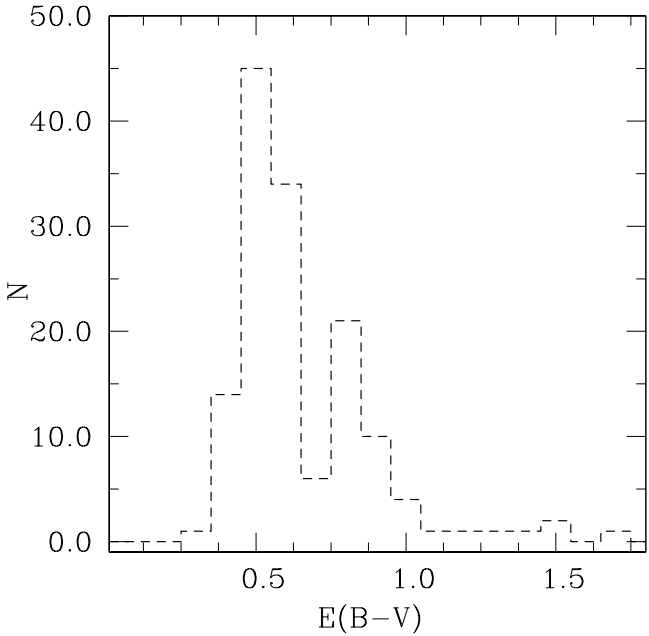,width=9cm,height=9cm}}
\caption{Individual reddening distribution for stars 
in the field of Trumpler~15 having $UBV$ photometry.}
\end{figure}

\begin{table*}
\tabcolsep 0.30cm
\caption{Photometric candidate members of the open cluster Trumpler~15 having
spectral classification. This is taken from Morrell et al (1988).
In the second column we list the numbering from Grubissich (1968).}
\begin{tabular}{cccccccccc}
\hline
\multicolumn{1}{c}{ID} &
\multicolumn{1}{c}{Grub.} &
\multicolumn{1}{c}{$Sp.~Type$} &
\multicolumn{1}{c}{$\alpha$}  &
\multicolumn{1}{c}{$\delta$}  &
\multicolumn{1}{c}{$V$} &
\multicolumn{1}{c}{$(B-V)$}  &
\multicolumn{1}{c}{$(U-B)$}  &
\multicolumn{1}{c}{$(V-R)$}  &
\multicolumn{1}{c}{$(R-I)$}  \\
\hline   
   2&   2&  O9-III &  10:42:47.08&   -59:05:29.50&     9.438&     0.213&    -0.816&     0.091&     0.168\\ 
   3&  15&  B0.5-IV&  10:42:44.70&   -59:07:14.80&    10.054&     0.224&    -0.743&     0.119&     0.210\\ 
   4&  26&  B1-V   &  10:42:31.80&   -59:04:18.00&    10.663&     0.224&    -0.714&     0.079&     0.144\\ 
   6&   7&  B2.5-V &  10:42:46.11&   -59:06:01.70&    10.582&     0.337&    -0.558&     0.198&     0.298\\  
   7&   3&  B2 -V  &  10:42:47.76&   -59:05:43.10&    10.627&     0.235&    -0.631&     0.119&     0.188\\ 
   8&  13&  B1-V   &  10:42:45.45&   -59:06:41.60&    10.861&     0.284&    -0.680&     0.149&     0.248\\ 
   9&   4&  B1-V   &  10:42:43.92&   -59:05:24.70&    11.019&     0.261&    -0.712&     0.127&     0.204\\
  10&  23&  B0-V   &  10:42:33.86&   -59:05:39.35&    11.059&     0.342&    -0.613&     0.111&     0.263\\  
  15&   5&  B5-V   &  10:42:45.06&   -59:05:45.60&    11.431&     0.304&    -0.590&     0.157&     0.267\\ 
  17&  10&  B2-V   &  10:42:49.77&   -59:06:05.20&    11.609&     0.316&    -0.470&     0.148&     0.230\\ 
  18&  18&  O9-I   &  10:42:39.82&   -59:08:32.20&    11.250&     1.286&    -0.252&     0.850&    -1.119\\ 
  25&   9&  B1-V   &  10:42:45.30&   -59:06:16.30&    12.636&     0.399&    -0.389&     0.224&     0.321\\ 
  26&  25&  B5-V   &  10:42:32.90&   -59:04:50.00&    12.682&     0.311&    -0.370&     0.141&     0.256\\ 
  28&  19&  09-V   &  10:42:32.70&   -59:07:40.00&    12.646&     0.316&    -0.369&     0.136&     0.305\\
  42&  22&  B9-V   &  10:42:40.80&   -59:06:18.70&    13.200&     0.431&    -0.055&     0.184&     0.315\\
  44&  21&  B0-III &  10:42:05:00&   -59:07:20.00&    13.242&     0.385&    -0.231&     0.180&     0.315\\  
\hline
\end{tabular}
\end{table*}

\noindent
The bulk of the stars are confined within a region defined
by two ZAMS shifted by $E(B-V)$~=~0.30 and 0.70 (dashed lines),
respectively.
The stars which occupy this region amount to 90 and
have a mean reddening $E(B-V)~=~0.52\pm0.07 (r.m.s.)$.\\
Since the spread in reddening is larger than the photometric errors,
which are typically $\Delta(U-B) \approx \Delta(B-V) \approx 0.04$,
this indicates the possible presence of differential reddening 
across the cluster.\\

There seems to be another population of stars more reddened,
which lies beyond the ZAMS shifted by $E(B-V)$~=~0.70 and  clearly
detaches from the previous group. These stars are indicated with
open squares, and have a mean reddening $E(B-V)~=~0.92\pm0.22 (r.m.s.)$.\\

Some additional indications in this sense derive also from Fig.~6,
where we plot the individual reddenings distribution function.
A clear peak is centered at $E(B-V) \approx$~0.50, while a
second well separated peak is visible, centered at $E(B-V) \approx~0.80-0.90$.
In any case there is a large spread in reddening, 
which suggests again the presence of differential
reddening across the cluster, as recently found also
by Yadav \& Sagar (2001).\\

\noindent
As a working hypothesis, we shall consider as candidate cluster members 
the stars 
having $E(B-V)$~=~0.52$\pm$0.07, and investigate how they distribute
in the reddening corrected CMDs.
In Fig.~7 we plot the reddening corrected CMDs in the planes
$V_o - (B-V)_o$ (left panel) and $V_o - (U-B)_o$ (right panel)
using the same symbols as in Fig.~5, and by provisionally
adopting $R_V~=~3.0$ (Feinstein et al 1980). 
All the stars brighter than $V_o < 11.5$ define a nice sequence.
There is just one exception, which is the star \#18 (Grubissich
numbering) which has  a very large reddening. The analysis  of its 
colours (see Table~2) seems to support the idea that this star
is actually a blend of two stars, a blue supergiant with a red 
companion (see also the discussion in Feinstein et al (1980).
Further sprectra are necessary to better clarify this issue.\\

Below $V_o$~=~11.5 stars having smaller reddening start
to mix with stars having larger reddening.
We interpret this fact in the sense that up to $V_o$~=~11.5 Trumpler~15
clearly emerges from the background, whereas below this magnitude
it smoothly merges with the general Galactic disk field.
This fact was not evident in previous photometry which
were not sufficiently deep.
A final check is to look at the spatial distribution 
of members and non-members in the field of the cluster.
This is presented in Fig.~8, where again filled circles
represent candidate members and open squares indicate candidate
non-members.
It seems that our selection criterion of members
and non-members worked fine, since member stars, as expected,
tend to concentrate in the cluster core, whereas non-members
preferentially distribute in the cluster outskirts.\\

In conclusions, we suggest that the 90 stars having $E(B-V)~=~0.52\pm0.07$
are probable members, whereas all the other stars having 
$E(B-V)~=~0.92\pm0.22$ are field stars. It is worth noticing here that
in a recent work DeGioia-Eastwood et al (2001) found that the field
stars in a region close to $\eta$~Carin\ae~ have indeed a reddening 
$E(B-V) \approx 1.0$.\\

Morrell et al (1988) provide spectral classification of 21 stars in the field
of Trumpler~15. We have combined the common stars (16) to obtain absolute
magnitudes and colors and derive another estimate of the cluster mean reddening.
These stars are listed in Table~2. We derive the reddening $E(B-V)$ by
adopting the intrinsic color indices of OB giants, supergiants and dwarf stars
in the $UBVRIJHKLM$ system provided by Wegner (1994).
For the 16 stars in Table~2 we obtain $E(B-V)~=~0.52\pm0.06$,
in perfect agreement with the mean reddening derived from the analysis of the
color-color diagram. 
This result makes us confident when using the photometric 
probable candidate members above derived.

\subsection{Differential reddening}
As already outlined, the large reddening spread visible in Figs~5 and 6
can be due to the presence of differential reddening across the cluster.
This possibility has been already discussed by Yadav and Sagar (2001)
by using data from the literature, and confirmed by the mid-infrared 
observations obtained by Smith et al (2000, fig.~2) with the $MSX$ satellite,
which show how the region of Trumpler~15 is unevenly obscured.\\
It is interesting to look for a possible
relation of the color excess of each candidate member with its position
on the plane of the sky, which would confirm the presence of patchy
absorption. To this aim, we plotted in Fig.~9 the distribution of colour 
excesses across the cluster. In this figure the size
of the points is proportional to the color excess. By inspective
this plot one can readily recognizes how the obscuration  toward
Trumpler~15 is really irregular.

\begin{figure}
\centerline{\psfig{file=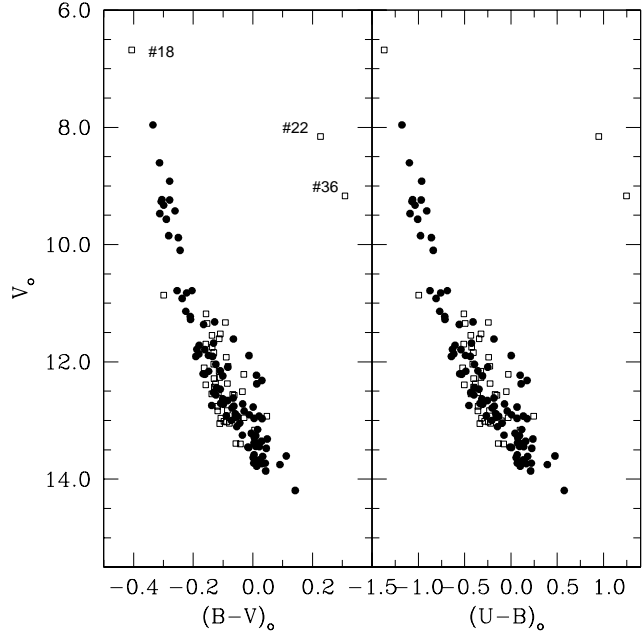,width=9cm,height=9cm}}
\caption{Reddening corrected CMDs for stars in the field of Trumpler~15
having $UBV$ photometry. Symbols are as in Fig.~5. Some interesting stars
have been labeled with their Grubissich number.}
\end{figure}

\begin{figure}
\centerline{\psfig{file=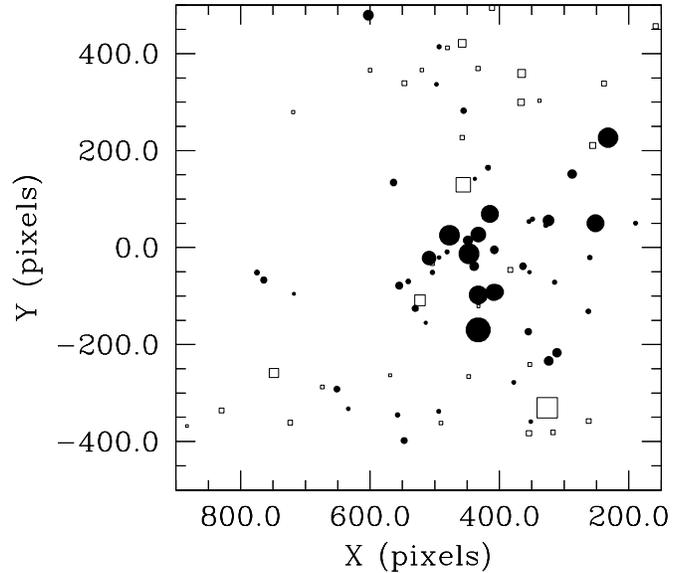,width=9cm,height=9cm}}
\caption{Spatial distribution of candidate members and non-members in 
the field of Trumpler~15
having $UBV$ photometry. Symbols are as in Fig.~5 and
the size of the stars is proportional to their
magnitudes.}
\end{figure}

\begin{figure}
\centerline{\psfig{file=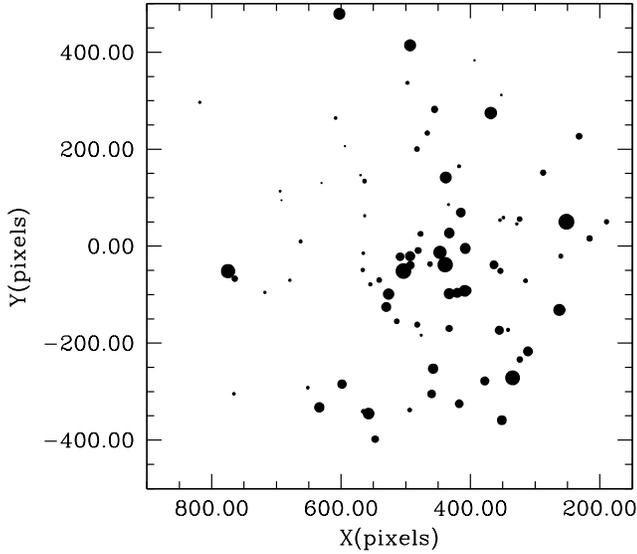,width=9cm,height=9cm}}
\caption{Spatial distribution of color excesses of 
the candidate member stars in the field of Trumpler~15. 
The size of each star is proportional to its color
excess.}
\end{figure}

\section{The extinction toward Trumpler~15}
There is some disagreement in the literature
whether the extinction toward this cluster is normal
($R_V$~=3.0, Feinstein et al 1980) or anomalous 
(larger $R_V$, Tapia et al 1988).\\
A recent estimate of $R_V$~=~4.0 is reported by Patriarchi et al (2001),
but is based on just one star (\# 18) which actually is 
a blend of two stars, as discussed above.\\
In this section we want to re-investigate
this problem by using our $UBVRI$ data combined
with near IR $JHKL$ data from Tapia et al (1988) and
spectral classifications from Morrell et al (1988).

\subsection{UBV data}
$UBV$ data can be used to derive an estimate of the $R_V$
parameter, defined as the ratio of the total to selective absorption
 $\frac{A_V}{E(B-V)}$, by means of  the variable extinction
method (see, for instance, Johnson 1966a).
This method requires also the knowledge of the absolute
magnitude and intrinsic colors of individual member stars.
We use the empirical ZAMS taken from Schmidt-Kaler (1982)
to derive absolute magnitude and intrinsic colors.\\
The result is shown in Fig.~10, where all the probable 
photometric members
selected in the previous Section  have been considered.
The least squares fit through the data provides  $R_V~=~2.89\pm0.28$.
\noindent 
The absolute distance modulus,
defined as the ordinate for E$(B-V)$~=~0, turns out to be
$(m-M)_0~=~11.90\pm0.25$, which implies a distance
from the Sun of $2.4\pm0.3$ kpc.\\
The reported errors are the results of the linear fit, obtained
by assuming that both the color excesses and the magnitueds have their
uncertainty.
However, due to the small range in color excess, the result we obtained 
has to be considered -with some caution- as no more than an indication. 

\begin{table*}
\tabcolsep 0.50cm
\caption{Extinction parameters of Trumpler~15 stars with near IR
photometry and spectral classification.}
\begin{tabular}{ccccccc}
\hline
\multicolumn{1}{c}{ID} &
\multicolumn{1}{c}{Grub.} &
\multicolumn{1}{c}{$Sp.~Type$} &
\multicolumn{1}{c}{$E(B-V)$}  &
\multicolumn{1}{c}{$A_V$}  &
\multicolumn{1}{c}{$R_V,(b)$}  &
\multicolumn{1}{c}{$R_V,(a)$} \\
\hline   
   2&   2&  O9-III &  0.473& 1.36$\pm$0.05&  2.87$\pm$0.13&  2.23\\ 
   3&  15&  B0.5-IV&  0.464& 1.29$\pm$0.13&  2.78$\pm$0.16&  2.92\\ 
   4&  26&  B1-V   &  0.454& 1.15$\pm$0.21&  2.53$\pm$0.29&  2.48\\ 
   6&   7&  B2.5-V &  0.532& 1.79$\pm$0.17&  3.36$\pm$0.25&  2.98\\  
   7&   3&  B2 -V  &  0.445& 1.50$\pm$0.33&  3.37$\pm$0.41&  2.96\\ 
   8&  13&  B1-V   &  0.514& 1.45$\pm$0.11&  2.82$\pm$0.20&  2.68\\ 
   9&   4&  B1-V   &  0.491& 1.25$\pm$0.09&  2.55$\pm$0.17&  2.37\\
  10&  23&  B0-V   &  0.602& 2.10$\pm$0.24&  3.49$\pm$0.37&  2.59\\  
  15&   5&  B5-V   &  0.454& 1.49$\pm$0.07&  3.28$\pm$0.15&  2.47\\ 
  17&  10&  B2-V   &  0.526& 1.45$\pm$0.13&  2.76$\pm$0.22&  2.61\\ 
  18&  18&  O9-I   &  1.556& 6.98$\pm$0.39&  4.48$\pm$0.65&  3.45\\ 
  25&   9&  B1-V   &  0.629& 2.42$\pm$0.19&  3.85$\pm$0.32&  3.19\\ 
  26&  25&  B5-V   &  0.461& 1.58$\pm$0.23&  3.42$\pm$0.40&  2.96\\ 
  28&  19&  09-V   &  0.596& 1.43$\pm$0.19&  2.40$\pm$0.35&  2.41\\
  42&  22&  B9-V   &  0.501& 1.70$\pm$0.16&  3.39$\pm$0.30&  2.83\\
  44&  21&  B0-III &  0.615& 2.30$\pm$0.31&  3.74$\pm$0.46&  3.19\\  
\hline
\end{tabular}
\end{table*}

\begin{figure}
\centerline{\psfig{file=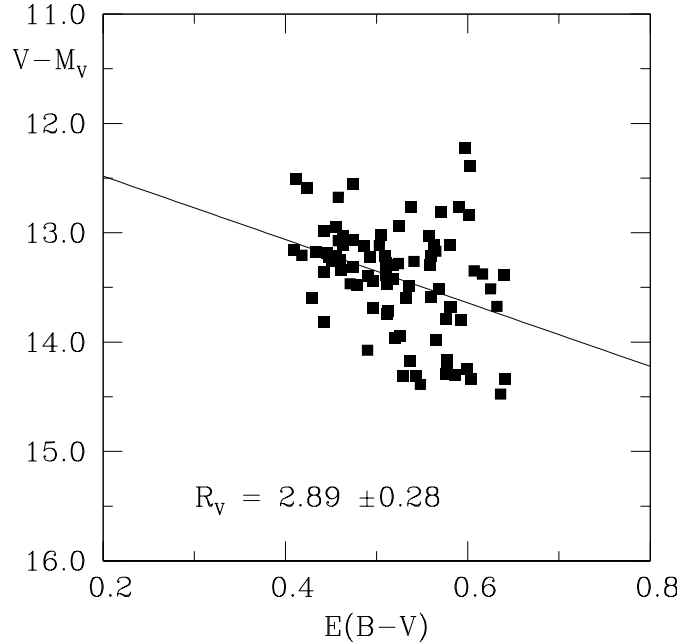,width=9cm,height=9cm}}
\caption{Differential reddening plot for all the MS stars
in the studied region. The solid line is a least squares fit
through the data.}
\end{figure}

\subsection{Near IR data}
To obtain another estimate of $R_V$ we combine
our $UBVRI$ photometry with the near IR photometry
in the $JHKL$ bands from Tapia et al (1988).
We have 30 stars in common, but we perform our analysis
only by considering the star having 
spectral classification.
Basically, there are two methods to derive and estimate of $R_V$
from near IR photometry.\\
The first method {\it (a)} is based on the following 
approximate relation

\begin{equation}
R_V \approx 1.1 \times \frac {E(V-K)}{E(B-V)}  ,
\end{equation}

\noindent
reported by Whittet (1990).
We use Wegner (1994) intrinsic colour indices to derive the
color excess  E$(B-V)$ and E$(V-K)$. The mean value for the stars
in Table~2 turns out to be

\[
R_V = 2.77 \pm 0.33 (r.m.s.)
\]

\noindent
in fine agreement with the variable extinction method.\\

The second method {\it (b)} makes use of the extinction curve, and was devised 
by Morbidelli et al (1997).
From the observed IR fluxes, the colour excesses E$(\lambda - V)$
have been found using the intrinsic colours by Wegner (1994).
The quantity $A_V$ has then been determined with a least squares
solution, by fitting (Cardelli et al, 1989):

\begin{equation}
E(\lambda -V) = A_V \times (R_L(\lambda) - 1)
\label{morbi}
\end{equation}

\noindent
where $\lambda = R,I,J,K,L$ and $R_L$ is the extinction curve
$\frac{A_\lambda}{A_V}$ taken from Rieke \& Lebofsky (1985).\\
From the derived $A_V$ and the known E$(B-V)$ we obtain
$R_V$.
Since the fitting equation~\ref{morbi} is a homogeneous one,
the uncertainty of each $A_V$ hase been computed by considering that we have 
$N-1$ degree of freedom ($N$ being the number of photometric
bands available). This implies that we can obtain only a lower limit
on the $R_V$ uncertainty, since it is not easy to take
into account spectral mis-classification and hence inaccuracy
in the adopted intrinsic colors (Patriarchi et al 2001).\\

The results are summarized in Table~3, where the identification
is reported together with the individual reddening E$(B-V)$,
the total absorption $A_V$, and the ratio 
of total-to-selective absorption $R_V~=~\frac{A_V}{E(B-V)}$, these latter
derived from the method {\it (b)}. The last column reports the value of $R_V$
derived from the method {\it (a)}.
A weighted mean yields $R_V~=~3.09\pm0.44$. in fine agreement both
with the results of method {\it (a)}, and with the finding of the
differential reddening method.
In fact the three values we obtain for $R_V$ clearly overlap,
and a weighted mean yields the final estimate

\[
R_V = 2.89\pm0.19  ,
\]

\noindent
which means that the interstellar reddening law toward Trumpler~15 appears
normal, as previously suggested for instance by Feinsten et al (1980).

\section{Age, distance and pre MS candidates}
We already obtained an estimate of Trumpler~15 distance from
the  analysis of the variable extinction plot in Section~6.
Here we address the issues of the distance and the age directly by
considering the reddening corrected CMDs for the member stars.
The correction to the magnitudes is done by computing
$V_o = V -3.0 \times E(B-V)$ for each star. This is plausible
since we have shown that the reddening law in the direction of the cluster
is normal.

\begin{figure}
\centerline{\psfig{file=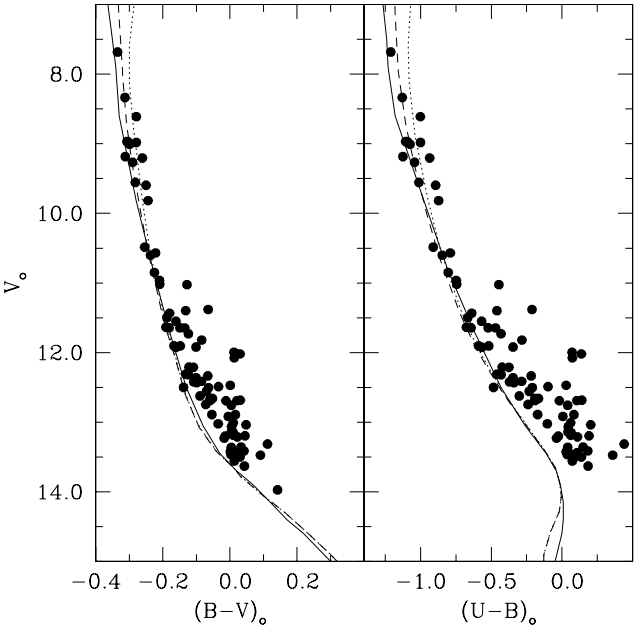,width=9cm,height=9cm}}
\caption{Reddening corrected CMD for the stars members
of Trumpler~15. The solid line is the empirical ZAMS form Schmidt-Kaler (1982)
Super-imposed are  post MS isochrones from Girardi et al 2000
(dashed and  dotted lines). See text for more details.}
\end{figure}

\subsection{Age and Distance}
The reddening corrected CMDs in the planes $V_o-(B-V)_o$
and $V_o-(U-B)_o$ are showed in the left and right panel of 
Fig.~11, respectively.
The solid line super-imposed 
is the Schmidt-Kaler (1982) ZAMS shifted by $(m-M)_o$~=~11.90,
which fits the blue edge of the stars distribution,
supporting the distance modulus estimate derived in
the previous Section.\\
The brightest stars depart somewhat
from the ZAMS, suggesting that they are evolving toward
the Red Giant region. This is confirmed by the location of the dashed and dotted lines,
which are a 2 (dashed line) and 6 (dotted line) 
million yrs solar metallicity isochrones taken from
Girardi et al (2000). 
The bright evolved stars lie between the two
isochrones implying that 
the age of this cluster is between 2 and 6 million yrs.

\subsection{A gap in the MS?}
Also in these diagrams a gap appears at $V_o \approx 10.5$, 
$(B-V)_o \approx -0.25$, which means that the gap is not
affected by the removal of non-members.\\ 
Notoriously, open star clusters often exhibit gaps in their MS
-see for instance the study of 
Mermilliod (1976) and Rachford \& Canterna (2000).
It is possible to estimate the probability that a lack
of stars in a mass interval is due to random
processes by computing

\begin{equation}
P_{gap} = (\frac{m_{sup}}{m_{inf}})^{(-N \times x)}
\end{equation}

\noindent
where $m_{sup}$ and $m_{inf}$ are the border masses of the gap,
$m_{sup} > m_{inf}$, $x$ is the exponent of the Initial Mass Function
(IMF) and $N$ the number of stars above the gap, i.e
with $m > m_{sup}$ (see Scalo (1986) and Giorgi et al (2001) for details).
We derive star masses from the reddening corrected colors and magnitudes,
finding that $m_{sup} \approx 9 M_{\odot}$ 
and $m_{inf} \approx 7 M_{\odot}$ (Girardi et al 2000).\\
The number of stars brighter than the gap is $N~=~12$.
By assuming a Salpeter (1955) IMF, with $x~=~1.35$ we yield
 $P_{gap}~=~8 \times 10^{-4}$.
Since this value is very small, we therefore 
conclude that the gap is a real feature.\\

The spectral type at the borders of the gap can be inferred from
the absolute magnitudes and colours, which we derive in this case from
Johnson (1966b), since
Wegner (1994) does not report the absolute $(U-B)$ color.  
The upper border is defined by $M_V=-2.00$, $(B-V)_0=-0.25$,
$(U-B)_0=-0.90$, which corresponds to spectral type $B1$; the lower
border on the other hand is defined by
$M_V=-1.00$, $(B-V)_0=-0.15$, $(U-B)_0=-0.52$, which corresponds 
to spectral type $B5$.
We therefore conclude that the gap is the $B1-B5$ gap, already found by
other authors (Newell 1973), which could not be detected in previous
photometric studies.

\begin{figure}
\centerline{\psfig{file=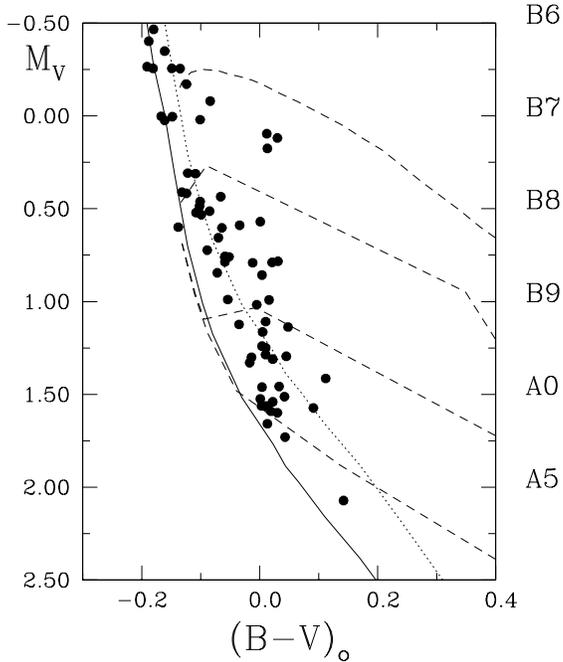,width=9cm,height=9cm}}
\caption{A zoom of the lower part of the reddening corrected CMDs.
The solid line is a post MS isochrone for the age of 6 million yrs,
the dotted line in the left panel 
is the same isochrone, but shifted by 0.70 to mimic
the locus of unresolved binaries, and the
dashed lines are from the bottom to the top
pre MS isochrones for the ages of 2, 3, 5 and 7 million yrs.}
\end{figure}

\subsection{Pre MS candidates.}
In Fig.~12 we propose a zoom of the lower part of the MS 
of the reddening corrected CMDs in the planes $V_o-(B-V)_o$.
On the right we list the spectral type of the stars derived
from the intrinsic magnitudes and colors (Johnson 1966b).
Our aim is to compare the distribution of stars with pre MS isochrones
to see whether the position of these stars is compatible with the possibility
that they are in contracting phase toward the ZAMS.
To this scope, we have super-imposed a post MS isochrone (solid line)
for the age of 6 million yrs, taken from Girardi et al (2000).
In addition, we have overlaid three pre MS isochrones (dashed lines) taken from
D'Antona \& Mazzitelli (1994), and kindly provided by Paolo Ventura.
From the bottom to the top they have ages of 2, 3, 5  and 7 million yrs.\\
First of all we would like to note that the width of the MS here is much larger
than photometric errors, which at these magnitude levels amount at
$\Delta (B-V) \approx 0.04$ mag in color.
The possibility that these stars are all binaries
is ruled out as well, since the unresolved binaries
isochrone, which lies only 0.70 mag above the 6 million yrs isochrone, encompasses
only a fraction of these stars. To guide the eye we have drawn
an 6 million yrs unresolved binaries
isochrone as a dotted line in Fig.~12.\\
Finally, it is obvious that some of these stars can be field stars that
we were not able to remove on the base of the sole reddening.
However
the mean reddening of the stars we have discarded 
(see Section~5) is 
E$(B-V) \approx 0.9$, very similar to that 
found by DeGioia-Eastwood et al (2001) for their background field.\\
This makes us confident on the reliability of the reddening 
criterion to disentangle cluster stars from field stars
(see also Cudworth et al 1993 for a discussion on this subject).\\

In addition DeGioia-Eastwood et al (2001)  find that
Trumpler~14 and 16, two very well known clusters located basically at the same 
distance from the Sun as Trumpler~15 and probably coeval, 
exhibit pre MS stars starting from 
$M_V \approx 0.00$ ($Log \frac{L}{L_{\odot}} \approx 2.5$).\\
A nice feature in the CMD of Fig.~4 is the presence of a possible 
{\it turn-on} at 
$V \approx 14.5$, which corresponds to  $(M_V \approx +1.00)$.
These facts, together with the nice fit provided by 
pre  MS isochrones 
suggests that indeed some of the stars with spectral types later than
$B7-B8$ can be in contracting phase.
This would imply that a significant age spread is present in this cluster.\\

\noindent
Deep near-infrared photometry and spectroscopy can help to cast more
light on the nature of this population.

\begin{figure}
\centerline{\psfig{file=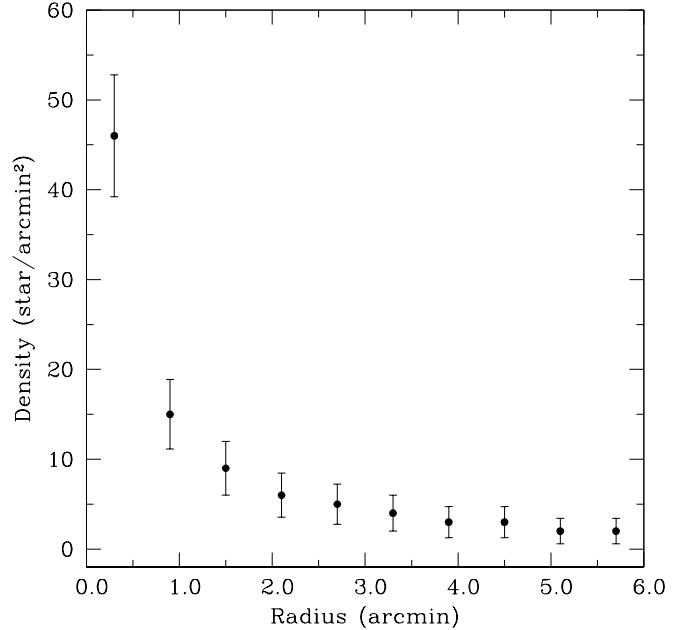,width=9cm,height=9cm}}
\caption{Density of stars in Trumpler~15 as a function of radius.}
\end{figure}

\section{Cluster size}
Trumpler~15 appears very compact on sky maps (see also Fig.~1).
According to WEBDA catalog, the cluster diameter is about
$14^{\prime}$, much larger than the region 
we covered.\\
We derived the surface stellar density by performing star counts
in concentring rings around stars \#6 (selected as cluster center)
and then dividing by their 
respective surfaces. The final density profile and the corresponding
poissonian
error bars are depicted in Fig.~13.
The surface density decreases sharply  up to $\approx 2^{\prime}$, then
the decrease continues very smoothly.
We can consider this radius as the core radius of the cluster.
Inside this radius the cluster dominates stars counts, outside 
it smoothly merges with the field.
A larger field coverage is necessary to give an estimate of the cluster radius,
since members seem to be evenly present in the field we covered.

\section{Conclusions}
In this paper we have presented new $UBVRI$ CCD photometry for Trumpler~15,
a young open cluster located in the Carina
spiral feature.\\
We identify  90 photometric members on the base of individual reddenings,
position on the CMDs and spatial distribution in the field.\\
Basing on this large sample we provide updated estimates of 
cluster fundamental parameters.
We find that the cluster is young, with an age between 2 and 6 million yrs,
contains a possible  population of pre MS candidates, which
deserves further investigation, and shows a gap in the MS at $V_o \approx
10.5$, that we suggest to be a real feature (the $B1-B5$ gap already found
in other clusters).\\
We place the cluster at $2.4\pm0.3$ kpc from the Sun.
Moreover we obtain  E$(B-V)$~=~0.52$\pm$0.07,  
and find that the extinction toward Trumpler~15 
can be considered normal.
Finally we estimate that the cluster has a core radius of 
about $2^{\prime}$.\\

Trumpler~15 appears to be located somewhat closer
to the Sun than Trumpler~14, Trumpler~16 and Collinder~232.
Nevertheless, the data suggest that Trumpler~15 
might belong to the complex
defined  by Trumpler~14, Trumpler~16, Collinder~232 and Collinder~228,
since it shares with these clusters the same age and the presence of
pre MS candidates. Moreover the extinction law seems to be
basically normal (with some local fluctuations) in the entire region.\\

\noindent
The most appealing scenario one can envisage 
is that all these clusters probably
formed together in the same recent Star Formation event.

\section*{Acknowledgments}
The author expresses his gratitude to ESO for kind hospitality,
and thanks Guido Barbaro, Mario Perinotto and Nando Patat 
for many fruitful conversations.
This study made use of Simbad and WEBDA.

\label{lastpage}

\end{document}